\documentclass[10pt, a4paper]{article}
\usepackage{lrec2022} 
\usepackage{multibib}
\newcites{languageresource}{Language Resources}
\usepackage{graphicx}
\usepackage{tabularx}
\usepackage{soul}
\usepackage{mathtools}
\usepackage{titlesec}
\titleformat{\section}{\normalfont\large\bfseries\center}{\thesection.}{1em}{}
\titleformat{\subsection}{\normalfont\SmallTitleFont\bfseries\raggedright}{\thesubsection.}{1em}{}
\titleformat{\subsubsection}{\normalfont\normalsize\bfseries\raggedright}{\thesubsubsection.}{1em}{}
\renewcommand\thesection{\arabic{section}}
\renewcommand\thesubsection{\thesection.\arabic{subsection}}
\renewcommand\thesubsubsection{\thesubsection.\arabic{subsubsection}}

\usepackage{epstopdf}
\usepackage[utf8]{inputenc}

\usepackage{hyperref}
\usepackage{xstring}

\usepackage{color}

\title{Privacy-Preserving Graph Convolutional Networks for Text Classification}

\name{Timour Igamberdiev, Ivan Habernal}

\address{Trustworthy Human Language Technologies \\
	Department of Computer Science \\
    Technical University of Darmstadt \\
	\texttt{\{timour.igamberdiev,ivan.habernal\}@tu-darmstadt.de} \\
	\texttt{www.trusthlt.org}
}

\abstract{
Graph convolutional networks (GCNs) are a powerful architecture for representation learning on documents that naturally occur as graphs, e.g., citation or social networks. However, sensitive personal information, such as documents with people's profiles or relationships as edges, are prone to privacy leaks, as the trained model might reveal the original input. Although differential privacy (DP) offers a well-founded privacy-preserving framework, GCNs pose theoretical and practical challenges due to their training specifics. We address these challenges by adapting differentially-private gradient-based training to GCNs and conduct experiments using two optimizers on five NLP datasets in two languages. We propose a simple yet efficient method based on random graph splits that not only improves the baseline privacy bounds by a factor of 2.7 while retaining competitive $F_1$ scores, but also provides strong privacy guarantees of $\varepsilon = 1.0$. We show that, under certain modeling choices, privacy-preserving GCNs perform up to 90\% of their non-private variants, while formally guaranteeing strong privacy measures.
\\ \newline  }

\usepackage{microtype}
\usepackage{amsfonts}
\usepackage{amsthm}

\usepackage{booktabs}

\begin{document}

    \onecolumn
    \noindent \textbf{Privacy-Preserving Graph Convolutional Networks for Text Classification}

    \medskip
    \noindent Timour Igamberdiev and Ivan Habernal

    \bigskip
    This is a \textbf{camera-ready version} of the article accepted for publication at the \emph{13th Language Resources and Evaluation Conference (LREC 2022)}. The final official version will be published on the ACL Anthology website later in 2022: \url{https://aclanthology.org/}

    \medskip
    Please cite this pre-print version as follows.
    \medskip

\begin{verbatim}
@InProceedings{Igamberdiev.2022.LREC,
    title = {{Privacy-Preserving Graph Convolutional Networks
              for Text Classification}},
    author = {Igamberdiev, Timour and Habernal, Ivan},
    publisher = {European Language Resources Association},
    booktitle = {Proceedings of the 13th Language Resources
                 and Evaluation Conference},
    pages = {(to appear)},
    year = {2022},
    address = {Marseille, France}
}
\end{verbatim}
    \twocolumn

\maketitleabstract

\section{Introduction}
Many text classification tasks naturally occur in the form of graphs where nodes represent text documents and edges are task specific, such as articles citing each other or health records belonging to the same patient. When learning node representations and predicting their categories, models benefit from exploiting information from the neighborhood of each node, as shown in graph neural networks, and graph convolutional networks (GCNs) in particular \cite{Kipf.Welling.2017.ICLR}, making them superior to other models \cite{Xu.et.al.2019.ICLR,DeCao.et.al.2019.NAACL}.

While GCNs are powerful for a variety of NLP problems, like other neural models they are prone to privacy attacks. Adversaries with extensive background knowledge and computational power might reveal sensitive information about the training data from the model, such as reconstructing information about the original classes of a model \cite{hitaj2017deep} or even auditing membership of an individual's data in a model \cite{song2019auditing}. In order to preserve privacy for graph NLP data, models have to protect both the textual nodes and the graph structure, as both sources carry potentially sensitive information.

Privacy-preserving techniques, such as differential privacy (DP) \cite{Dwork.Roth.2013}, prevent information leaks by adding `just enough' noise during model training while attaining acceptable performance.
Recent approaches to DP in neural models attempt to trade off between noise and utility, with differentially private stochastic gradient descent (SGD-DP) \cite{Abadi.et.al.2016.SIGSAC} being a prominent example.
However, SGD-DP's design expects i.i.d.\ data examples to form
batches and `lots',
therefore its suitability for graph neural networks remains an open question.

In this work, we propose a methodology for applying differentially private stochastic gradient descent and its variants to GCNs, allowing to maintain strict privacy guarantees and performance.
Our approach consists of applying an easy-to-implement graph splitting algorithm to GCNs in the DP setting, partitioning a graph into
subgraphs
while avoiding additional queries on the original data. We adapt SGD-DP \cite{Abadi.et.al.2016.SIGSAC} to GCNs as well as propose a differentially-private version of Adam \cite{kingma2017adam}, Adam-DP.
We hypothesize that Adam's advantages, i.e.\ fewer training epochs, would lead to a better privacy/utility trade-off as opposed to SGD-DP.

We conduct experiments on five datasets in two languages (English and Slovak) covering a variety of NLP tasks, including research article classification in citation networks, Reddit post classification, and user interest classification in social networks, where the latter ones inherently carry potentially sensitive information calling for privacy-preserving models.
Our main contributions are twofold.
First, we show that DP training can be applied to the case of GCNs, with graph splitting and proper optimization recovering a lot of the dropped performance due to DP noise.
Second, we show that more sophisticated text representations further mitigate the performance drop due to DP noise, resulting in a relative performance of 90\% of the non-private variant, while keeping strict privacy ($\varepsilon = 1.0$ when using graph splits).
To the best of our knowledge, this is the first study that brings differentially private gradient-based training to graph neural networks.\footnote{Code available at
\url{https://github.com/trusthlt/privacy-preserving-gcn}
}

\section{Theoretical background in DP}
\label{sec:dp}

As DP does not belong to the mainstream methods in NLP, here we shortly outline the principles and present the basic terminology from the NLP perspective. Foundations can be found in \cite{Dwork.Roth.2013,Desfontaines.Pejo.2020}.

The main idea of DP is that if we query a database of $N$ individuals, the result of the query will be almost indistinguishable from the result of querying a database of $N -1$ individuals, thus 
preserving
each single individual's privacy to a certain degree. The difference of results obtained from querying any two databases that differ in one individual has a probabilistic interpretation. 

Dataset $D$ consists of $|D|$ documents where each document is associated with an individual whose privacy we want to preserve. A document can be any arbitrary natural language text, such as a letter, medical record, tweet, personal plain text passwords, or a paper review. Let $D'$ differ from $D$ by one document, so either $|D'| = |D| \pm 1$, or $|D'| = |D|$ with $i$-th document replaced. $D$ and $D'$ are called \emph{neighboring} datasets.

Let $A: D \mapsto y \in \mathbb{R}$ be a randomized function applied to a dataset $D$; for example a function returning the average document length or the number of documents in the dataset. This function is also called a \emph{query} which is not to be confused with queries in NLP, such as search queries.\footnote{In general, the query output is multidimensional $\mathbb{R}^k$; here we keep it scalar for the sake of simplicity.} In DP, this query function is a continuous random variable associated with a probability density $p_t(A(D) = y)$. Once the function $A(D)$ is applied on the dataset $D$, the result is a single draw from this probability distribution. This process is also known as a \emph{randomized algorithm}.
For example, a randomized algorithm for the average document length can be a Laplace density such that
$p_t(A(D) = y) = \frac{1}{2b}\exp \left( - \frac{|\mu - y|}{b} \right),$
where $\mu$ is the true average document length and $b$ is the scale (the `noisiness' parameter). By applying this query to $D$, we obtain $y \in \mathbb{R}$, a single draw from this distribution.

Now we can formalize the backbone idea of DP. Having two neighboring datasets $D$, $D'$, \emph{privacy loss} is defined as

\begin{equation}
\label{eq:privacy-loss1}
\ln \frac{p(A(D) = y)}{p(A(D') = y)}  .
\end{equation}

DP bounds this privacy loss by design. Given $\varepsilon \in \mathbb{R}: \varepsilon \geq 0$ (the \emph{privacy budget} hy\-per-pa\-ra\-me\-ter), all values of $y$, and all neighboring datasets $D$ and $D'$, we must ensure that

\begin{equation}
\label{eq:dp-def1}
\max_{\forall y}  \left| \ln \frac{p(A(D) = y)}{p(A(D') = y)} \right| \leq \varepsilon \: .
\end{equation}

In other words, the allowed privacy loss of any two neighboring datasets is upper-bounded by $\varepsilon$, also denoted as $(\varepsilon, 0)$-DP.\footnote{
$(\varepsilon, 0)$-DP is a simplification of more general $(\varepsilon, \delta)$-DP where $\delta$ is a negligible constant allowing relaxation of the privacy bounds \cite[p.~18]{Dwork.Roth.2013}.} The privacy budget $\varepsilon$ controls the amount of preserved privacy. If  $\varepsilon \rightarrow 0$, the query outputs of any two datasets become indistinguishable, which guarantees almost perfect privacy but provides very little utility. Similarly, higher $\varepsilon$ values provide less privacy but better utility. Finding the sweet spot is thus the main challenge in determining the privacy budget for a particular application \cite{Lee.Clifton.2011.ISC,Hsu.et.al.2014.CSFS}. An important feature of $(\varepsilon, \delta)$-DP is that once we obtain the result $y$ of the query $A(D) = y$, any further computations with $y$
cannot weaken the privacy guaranteed by $\varepsilon$ and $\delta$.

The desired behavior of the randomized algorithm is therefore adding as little noise as possible to maximize utility while keeping the privacy guarantees given by Eq.~\ref{eq:dp-def1}. The amount of noise is determined for each particular setup by the \emph{sensitivity} of the query  $\Delta A$, such that for any neighboring datasets $D, D'$ we have

\begin{equation}
\Delta A = \max_{\forall D, D'} \left( |A(D) - A(D')| \right) \: .
\end{equation}

The sensitivity corresponds to the `worst case' range of a particular query $A$, i.e., what is the maximum impact of changing one individual. The larger the sensitivity, the more noise must be added to fulfill the privacy requirements of $\varepsilon$ (Eq.~\ref{eq:dp-def1}). For example, in order to be $(\varepsilon, 0)$-DP, the Laplace mechanism must add noise $b = (\Delta A)^{-1}$ \cite[p.~32]{Dwork.Roth.2013}.
As the query sensitivity directly influences the required amount of noise, it is desirable to design queries with low sensitivity.

The so far described mechanisms consider a scenario when we apply the query only once. To ensure $(\varepsilon, \delta)$-DP with multiple queries\footnote{Queries might be different, for example querying the average document length first and then querying the number of documents in the dataset.} on the same datasets, proportionally more noise has to be added.

\section{Related work}

A wide range of NLP tasks have been utilizing \textbf{graph neural networks} (GNNs), specifically graph convolutional networks (GCNs), including text summarization \cite{xu2020discourse}, machine translation \cite{marcheggiani2018exploiting} and semantic role labeling \cite{zheng2020srlgrn}. Recent end-to-end approaches combine pre-trained transformer models with GNNs to learn graph representations for syntactic trees \cite{sachan2020syntax}.
\newcite{Rahimi.et.al.2018.ACL} demonstrated the strength of GCNs on predicting geo-location of Twitter users where nodes are represented by users' tweets and edges by social connections, i.e.\ mentions of other Twitter users. However, for protecting user-level privacy, the overall social graph has to be taken into account.

Several recent works in the NLP area deal with \textbf{privacy} using arbitrary definitions. \newcite{Li.et.al.2018.ACLShort} propose an adversarial-based approach to learning latent text representation for sentiment analysis and POS tagging. Although their privacy-preserving model performs on par with non-private models, they admit the lack of formal privacy guarantees. Similarly, \newcite{Coavoux.et.al.2018.EMNLP} train an adversarial model to predict private information on sentiment analysis and topic classification. The adversary's model performance served as a proxy for privacy strength but, despite its strengths, comes with no formal privacy guarantees. Similar potential privacy weaknesses can be found in a recent work by \newcite{Abdalla.et.al.2020.JAMIA} who replaced personal health information by semantically similar words while keeping acceptable accuracy of downstream classification tasks.

\newcite{Abadi.et.al.2016.SIGSAC} pioneered the connection of DP and deep learning by proposing SGD-DP that bounds the query sensitivity using gradient clipping and formally guarantees the overall privacy budget.
While originally tested on image recognition, they inspired subsequent work in language modeling using LSTMs \cite{McMahan.et.al.2018.ICLR}.
However, to the best of our knowledge, training
graph-based architectures with SGD-DP has not yet been explored. 
Two recent approaches utilize \emph{local DP}, that is adding noise to each node before passing it to graph model training \cite{Sajadmanesh.Gatica-Perez.2020.arXiv,Lyu.et.al.2020.SIGIR}, yet it is unclear whether it prevents leaking knowledge about edges. Our setup is different as we have access to the full dataset and preserve privacy of the entire graph.

As GCNs typically treat the entire graph as a single training example, \newcite{chiang2019cluster} proposed a more efficient training using mini-batching methods. Despite the NP-hardness of the general graph splitting problem \cite{bui1992finding}, they experimented with random partitioning and other clustering methods that take advantage of the graph structure \cite{karypis1998fast}.
It remains an open question whether splitting the graph into disjoint i.i.d.\ examples would positively affect our DP-approach where mini-batches parametrize the required amount of noise.

\section{Models}

\subsection{GCN as the underlying architecture}

We employ the GCN architecture \cite{Kipf.Welling.2017.ICLR} for enabling DP in the domain of graph-based NLP. GCN is a common and simpler variant to more complex types of GNNs which allows us to focus primarily on a comparison of the DP and non-DP models.

Let $\mathcal{G} = (\mathcal{V}, \mathcal{E})$ model our graph data where each node $v_i \in \mathcal{V}$ contains a feature vector of dimensionality $d$. GCN aims to learn a node representation by integrating information from each node's neighborhood. The features of each neighboring node of $v_i$ pass through a `message passing function' (usually a transformation by a weight matrix $\Phi$) and are then aggregated and combined with the current state of the node $h_i^l$ to form the next state $h_i^{l+1}$. Edges are represented using an adjacency matrix $A \in \mathbb{R}^{n \times n}$, where $n$ is the number of nodes in the graph. $A$ is then multiplied by the matrix $H \in \mathbb{R}^{n \times f}$, $f$ being the hidden dimension, as well as the weight matrix $\Phi$ responsible for message passing, learned during training. Additional tweaks by \newcite{Kipf.Welling.2017.ICLR} include adding the identity matrix to $A$ to include self-loops in the computation $\hat{A} = A + \mathbf{I}$, as well as normalizing matrix $A$ by the degree matrix $D$, specifically using a symmetric normalization $D^{-\frac{1}{2}}AD^{-\frac{1}{2}}$. This results in the following equation for calculating the next state of the GCN for a given layer $l$, passing through a non-linearity function $\sigma$:

\begin{equation}
H^{l+1} = \sigma \left( \hat{D}^{-\frac{1}{2}}\hat{A}\hat{D}^{-\frac{1}{2}}H^{(l)}\Phi^{(l)} \right)
\label{eq:gcn-layer}
\end{equation}
The final layer states for each node are then used for node-level classification, given output labels.

\subsection{Baseline model: SGD-DP}
\label{sec:sgd-dp.adam-dp}

SGD-DP \cite{Abadi.et.al.2016.SIGSAC} modifies the standard stochastic gradient descent algorithm to be differentially private. The DP `query' is the gradient computation at time step $t$: $g_t(x_i) \leftarrow \nabla_{\theta_t} \mathcal{L}(\theta_t, x_i)$, for each $i$ in the training set. To ensure DP, the output of this query is distorted by random noise proportional to the sensitivity of the query, which is the range of values that the gradient can take. As gradient range is unconstrained, possibly leading to extremely large noise, \newcite{Abadi.et.al.2016.SIGSAC} clip the gradient vector by its $\ell_2$ norm, replacing each vector $g$ with $\bar{g} = g / \max(1, \frac{||g||_2}{C})$, $C$ being the clipping threshold.
This clipped gradient is altered by a draw from a Gaussian: $\bar{g}_t(x_i) + \mathcal{N}(0, \sigma^2 C^2 I)$.

Instead of running this process on individual examples, \newcite{Abadi.et.al.2016.SIGSAC} actually break up the training set into `lots' of size $L$, being a slightly separate concept from that of `batches'. Whereas the gradient computation is performed in batches, SGD-DP groups several batches together into lots for the DP calculation itself, which consists of adding noise, taking the average over a lot and performing the descent $\theta_{t+1} \leftarrow \theta_t - \eta_t \tilde{\mathbf{g}}_t$.

Incorporating this concept, we obtain the overall core mechanism of SGD-DP:
\begin{equation}
	\tilde{\mathbf{g}}_t = \frac{1}{L} \left( \sum_{i \in L} \frac{\mathbf{g}_t(x_i)}{\max \left( 1, \frac{||\mathbf{g}_t(x_i)||_2}{C}  \right) } + \mathcal{N}(0, \sigma^2 C^2 \mathbf{I}) \right)
	\label{eq:dp-sgd-def}
\end{equation}

\subsection{Our DP extension: Differentially-private Adam}

In this paper, we also propose a DP version of Adam \cite{kingma2017adam}, a widely-used default optimizer in NLP \cite{ruder2016overview}. As Adam shares the core principle of gradient computing within SGD, to make it differentialy private we add noise to the gradient following Eq.~\ref{eq:dp-sgd-def}, prior to Adam's moment estimates and parameter update. Adam-DP thus guarantees privacy like SGD-DP does, namely (1) by DP privatizing the query, that is the gradient, and (2) by a  composition theorem, that is a sequence of DP mechanisms remains DP.

Despite their conceptual simplicity, both SGD-DP and Adam-DP have to determine the amount of noise to guarantee $(\varepsilon, \delta)$ privacy. \newcite{Abadi.et.al.2016.SIGSAC} proposed the moments accountant which we present in detail
in Appendix \ref{sec:moments.accountant}.

\subsection{Our approach: Graph cuts for improved DP performance}
\label{sec:graph.splitting}

We propose a simple yet effective treatment of the discrepancy between GCN training (that is, taking the entire graph as a single example to maximally utilize the contextual information of each node) and DP-version of SGD and Adam (which requires a set of i.i.d.\ examples to form batches and `lots' in order to distribute DP noise effectively).

The unit for which our method provides a DP guarantee is a full graph, including all of its nodes and edges, which contrasts with other notions of DP for graphs such as Edge DP and Node DP \cite{kasiviswanathan2013analyzing}, which only protect edges, or nodes with all of their adjacent edges, respectively. As DP operates with the notion of `neighboring datasets' \cite[Sec.~4]{Desfontaines.Pejo.2020},
training a GCN privately on the full graph means that \emph{any other graph} is neighboring. It also implies that each individual's privacy in that graph is protected, which is \emph{the} goal of differential privacy. The other extreme would be to completely ignore the graph structure and train GCN on individual nodes; using DP, it would again protect each individual's privacy, but any advantage of graph structure would be ignored.

We thus propose a sweet-spot approach, that is splitting the graph into disconnected subgraphs. We experiment with different numbers of subgraphs to find the best trade-off. In order to avoid any further dataset queries that might
require a larger privacy budget, we utilize random masking of the adjacency matrix so no additional DP mechanism is required.

Our algorithm
creates a random index tensor for all nodes in the training set, which is then split into $s$ groups, corresponding to the number of desired subgraphs.
If the number of nodes $n$ is divisible by $s$, then all subgraphs have equal sizes of nodes ($\frac{n}{s}$). If $n$ is not divisible by $s$, then $n\bmod{s}$ subgraphs have $\lfloor\frac{n}{s+1}\rfloor$ nodes, while the rest have $\lfloor\frac{n}{s}\rfloor$.
These indexes are then used to mask the original graph during training.
This step is performed once during data preprocessing and requires very little additional computational time or memory requirements.

\paragraph{Privacy guarantee of graph cuts}
We start by summarizing the main DP argument of SGD-DP \cite{Abadi.et.al.2016.SIGSAC}. In particular, any two gradient vectors are made `indistinguishable' from each other up to factor $\exp(\varepsilon)$ and summand $\delta$. Gradients are computed over mini-batches. This means that the presence or absence of an individual in the mini-batch is protected by SGD-DP. Furthermore, mini-batches are disjoint, such that each individual is associated with a unique mini-batch only.
The disjoint requirement stems from DP being defined through the notion of neighboring datasets, where a given individual's record in the dataset cannot appear in any of its neighboring datasets (see Section \ref{sec:dp}).
SGD-DP with mini-batches is $(\varepsilon, \delta)$-DP \cite{Abadi.et.al.2016.SIGSAC}.

In our graph scenario, we cut the graph into several disconnected subgraphs that are equivalent to `mini-batches'. When trained by SGD-DP, the gradients are again computed over each `mini-batch' (subgraph) and privatized.
Now having each individual (a single node and all its edges) in one `mini-batch', the presence or absence of that individual is again protected by DP as desired. Therefore SGD-DP on GNN with disjoint subgraphs is $(\varepsilon, \delta)$-DP.

Finally, our graph splitting algorithm is completely random. It does not query (in the DP sense) any information about the graph and its output is independent of a presence or absence of any individual. As such, the random graph splitting algorithm does not consume any privacy budget. Overall, this makes our approach $(\varepsilon, \delta)$-DP.

Our method is thus easy to implement, does not require much computational overhead and fits very well into the DP scenario. We extensively compare our results using different subgraph sizes in the non-private and private settings
in Section~\ref{sec:results}.

\section{Experiments}

\subsection{Datasets}
\label{sec:datasets}

We are interested in a text classification use-case where documents are connected via undirected edges, forming a graph. While structurally limiting, this definition covers a whole range of applications.
We perform experiments on five single-label multi-class classification tasks.
The \textbf{Cora}, \textbf{Citeseer}, and \textbf{PubMed} datasets \cite{Yang.et.al.2016.ICML,Sen.et.al.2008.AIMag,McCallum.et.al.2000.IR,Giles.et.al.1998.DL} are widely used citation networks
of research papers where citing a paper $i$ from paper $j$ creates an edge $i-j$. The task is to predict the category of the particular paper.

The \textbf{Reddit} dataset \cite{Hamilton.et.al.2017.NeurIPS} treats the `original post' as a graph node and connects two posts by an edge if any user commented on both posts. Given the large size of this dataset (230k nodes; all posts from Sept.~2014) causing severe computational challenges, we sub-sampled 10\% of posts (only few days of Sept.~2014). The gold label corresponds to one of the top Reddit communities to which the post belongs to.

Unlike the previous English datasets, the \textbf{Pokec} dataset \cite{takac2012data,snapnets} contains an anonymized social network in Slovak. Nodes represent users and edges their friendship relations. User-level information contains many attributes in natural language (e.g., `music', `perfect evening'). We set up the following binary task: Given the textual attributes, predict whether a user prefers dogs or cats.\footnote{We decided against user profiling, namely age prediction for ad targeting \cite{Perozzi.Skiena.2015.WWW}, for ethical reasons. Our task still serves well the demonstration purposes of text classification of social network data.} Pokec's personal information including friendship connections shows the importance of privacy-preserving methods to protect this potentially sensitive information.
For the preparation details see Appendix \ref{sec:pokec-preprocessing}.

\subsection{Experiment setup}

We operate with three bench-marking scenarios. \textbf{Experiment A} is vanilla GCN without DP: The aim is to train the GCN without any privacy mechanism, evaluating also influence on performance with less training data.
\textbf{Experiment B} is GCN with DP: We evaluate performance varying the amount of privacy budget as well as data size. We randomly sub-sample a certain percentage of nodes that then form a single training graph, as in standard GCN. The latter allows us to see the effects on performance of both adding noise and reducing training data.
\textbf{Experiment C} is GCN with graph splits: Evaluating performance varying the number of graph splits in the non-DP and DP settings.

\textbf{Implementation details.} As the $\delta$ privacy parameter is typically kept `cryptographically small' \cite{Dwork.Roth.2013} and, unlike the main privacy budget $\varepsilon$, has a limited impact on accuracy \cite[Fig.~4]{Abadi.et.al.2016.SIGSAC}, we fixed its value to $10^{-5}$ for all experiments. The clipping threshold is set at 1. We validated our PyTorch implementation by fully reproducing the MNIST results from \newcite{Abadi.et.al.2016.SIGSAC}.
We perform all experiments five times with different random seeds and report the mean and standard deviation. Early stopping is determined using the validation set.
See 
Appendix \ref{sec:hyperparams}
for more details on other hyperparameters.

\section{Results and analysis}
\label{sec:results}

\begin{table*}
	\begin{center}
		\begin{tabular}{lll|lll|cc}
			\multicolumn{3}{c}{\textbf{Non-DP}} & \multicolumn{3}{c}{\textbf{DP}} & \multicolumn{2}{c}{\textbf{DP split} } \\ 
			Maj. & SGD	& Adam & $\varepsilon$	&SGD 	&Adam &SGD & Adam	\\ \midrule
			\multicolumn{3}{l}{\textbf{CiteSeer}} &	1	&	-& -&0.35 &0.36	\\
			0.18 & 0.77 &0.79	& 2	&\textbf{0.36}	&\textbf{0.36} &0.35 &\textbf{0.36}	\\ \hline
			\multicolumn{3}{l}{\textbf{Cora}} 	& 1	&-	&- &0.55 &0.56	\\
			0.32 & 0.77&0.88		&2	&0.39	&0.52&0.55 &\textbf{0.57}	\\ \hline
			\multicolumn{3}{l}{\textbf{PubMed}} 	& 1	&-	& -&0.54 &0.52	\\
			0.40 & 0.49&0.79		&2	&0.38	&\textbf{0.54}&\textbf{0.54} &0.51	\\ \hline
			\multicolumn{3}{l}{\textbf{Pokec}} 	& 1	&-	&- &0.62 &0.72	\\
			0.50& 0.83&0.83	&2	&\textbf{0.75}	&0.66&0.64 &0.73	\\ \hline
			\multicolumn{3}{l}{\textbf{Reddit}} 	& 1	&-	&- &0.65 &0.79	\\
			0.15 & 0.68&0.88	&2	&0.46	&0.72&0.67 &\textbf{0.82}	\\ \hline
		\end{tabular}
	\end{center}
	\caption{\label{fig:experiments.ac} $F_1$ results for experiments A, B and C: full dataset without DP (first three columns, including a majority baseline), with DP and varying $\varepsilon$ (middle two columns), with DP using graph splits (right-most two columns). Best DP results are bold. Lower $\varepsilon$ corresponds to better privacy.}

\end{table*}

\begin{figure*}
	\includegraphics[width=\textwidth]{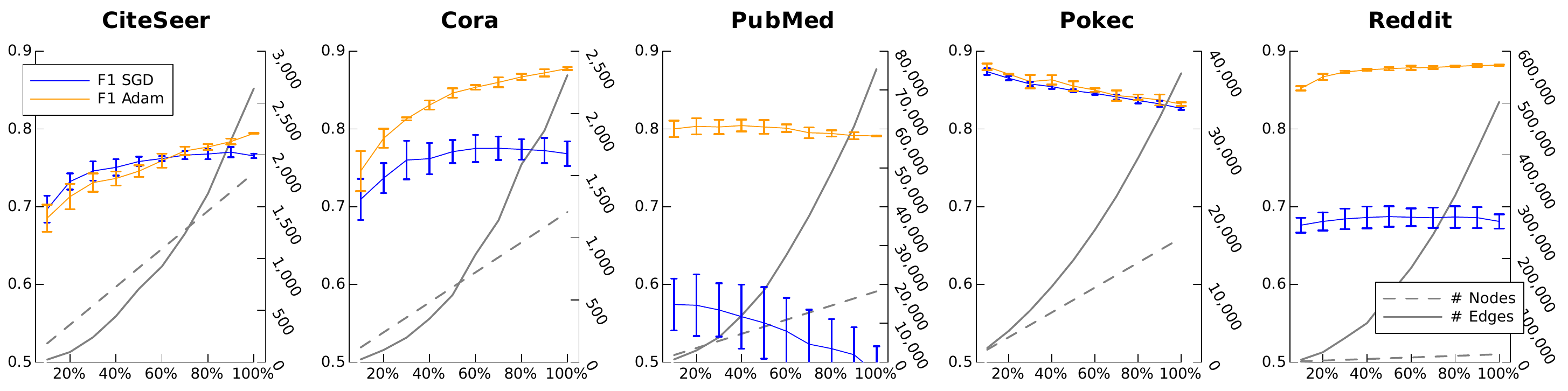}
	\caption{\label{fig:experiment.b} Experiment A: $F_1$ wrt.\ training data size (in \%), without DP.}
\end{figure*}

\subsection{Experiment A: Non-private GCN}

Table~\ref{fig:experiments.ac} shows the results on the left-hand side under `Non-DP'.
When trained with SGD, all datasets achieve fairly good results with the exception of PubMed, possibly due to PubMed having a much larger graph. The best of these is for Pokec, which could be due to its more expressive representations (BERT) and a simpler task (binary classification).

In comparison, in line with previous research \cite{ruder2016overview}, Adam outperforms SGD in all cases, with Pokec showing the smallest gap (0.826 and 0.832 for SGD and Adam, respectively).

Figure~\ref{fig:experiment.b} shows the non-DP results with increasing training data. We observe two contrasting patterns.
First, there is a clear improvement as training data increases (e.g. CiteSeer, with 0.70 F1 score at 10\% vs. 0.77 at 100\%).
Second, we observe the exact opposite pattern, with PubMed dropping from 0.57 at 10\% to 0.49 at 100\%, with a similar pattern for Pokec, or an early saturation effect for Reddit and Cora, where results do not increase beyond a certain point (at 20-30\% for Reddit with approximately 0.69 F1 score, 50\% for Cora at a score of 0.77).
We speculate that, with a larger training size, a vanilla GCN has a harder time to learn the more complex input representations. In particular, for PubMed and Pokec, the increasing number of training nodes only partially increases the graph degree, so the model fails to learn expressive node representations when limited information from the node's neighborhood is available. By contrast, Reddit graph degree grows much faster, thus advantaging GCNs.

\subsection{Experiment B: GCN with DP}
\label{sec:experiment-b}

The middle columns of Table~\ref{fig:experiments.ac} show results for
a privacy budget of $\varepsilon = 2.0$. As discussed further below, without splitting the graph into subgraphs, it is impossible to reach the lower $\varepsilon$ value of $1.0$, since computations become very unstable due to the large amount of noise. We do not report values larger than $\varepsilon = 2.0$
as their privacy protection diminishes exponentially.
We note four main patterns in this experiment.

\begin{figure*}[ht!]
	\includegraphics[width=\textwidth]{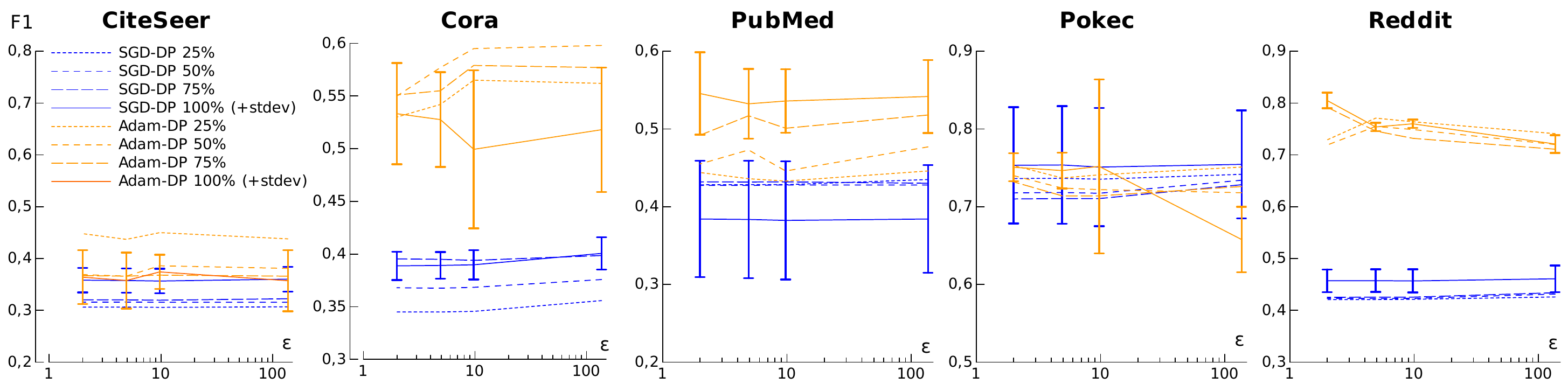}	
	\caption{\label{fig:experiment.d} Experiment B: $F_1$ with varying training data size (in \%) wrt.\ privacy budget $\varepsilon$, with DP.}
\end{figure*}

First, \textbf{SGD-DP results stay the same, regardless of the noise value added.}
This is quite unexpected, since higher added noise values would be anticipated to lead to lower results. One explanation for this pattern is that the gradients in vanilla SGD are already quite noisy, which may even help in generalization for the model, so the additional DP noise does not pose much difficulty beyond the initial drop in performance.

Second, \textbf{Adam-DP results outperform SGD-DP and can reach results close to the non-DP settings.}
It is worth noting that, when using default hyperparameters with a moderate learning rate of 0.01, Adam-DP results are very low, usually worse than SGD-DP. It is only when optimizing this learning rate that we see substantial improvements, with the best-performing learning rates being very large, in the case of Reddit as high as 100. In contrast, SGD-DP does not see much benefit from additional hyperparameter optimization.

Third, we see \textbf{bigger drops in performance in the DP setting for datasets with simpler input representations.}
Datasets of simpler input features can have results drop by more than half in comparison to the non-DP implementation.
In comparison to SGD-DP, Adam-DP is able to retain better performance even with the simpler input features for Cora and PubMed (e.g., drop $0.79 \to 0.54$ for PubMed).
Reddit and Pokec show the smallest drops from the non-DP to DP setting ($0.88 \to 0.72$ and $0.83 \to 0.66$ with Adam, for each dataset, respectively).
In fact, even for SGD-DP, Pokec results are very close to the non-DP counterpart ($0.83 \to 0.75$ F1 score).
Hence, Citeseer, Cora and PubMed, all using one-hot textual representations, show far greater drops in performance for the DP setting. The datasets utilizing GloVe (Reddit) and BERT (Pokec) representations perform far better.
Since this effect of feature complexity on DP performance is shown only through different datasets, we perform additional experiments on Pokec using BoWs, fastText \cite{grave2018learning} and BERT features for a proper `apples-to-apples' comparison
described in Section~\ref{sec:pokec-feature-comparison}.

\paragraph{Learning Curves with DP}
Figure~\ref{fig:experiment.d} shows the DP results both for varying $\varepsilon$ and with different training sub-samples (25\%, 50\%, 75\% and the full 100\%).
First, generally observed patterns are not the same for the learning curves in the non-DP setup (Experiment A). For instance, Adam exhibits the opposite pattern, e.g. Citeseer and Cora increase with more data for Adam without DP, but decrease for Adam-DP.

Second, we can see that \textbf{increasing the amount of data does not necessarily help in the DP setting}. For instance, while there is an improvement for SGD-DP with the Citeseer, Cora and Reddit datasets, results mostly get worse for Adam-DP, with the exception of PubMed.
Hence, increasing training data generally does not act as a solution to the general drop in performance introduced by DP.

\subsection{Experiment C: Graph splitting approach}

\begin{figure*}
    \centering
	\includegraphics[width=\linewidth]{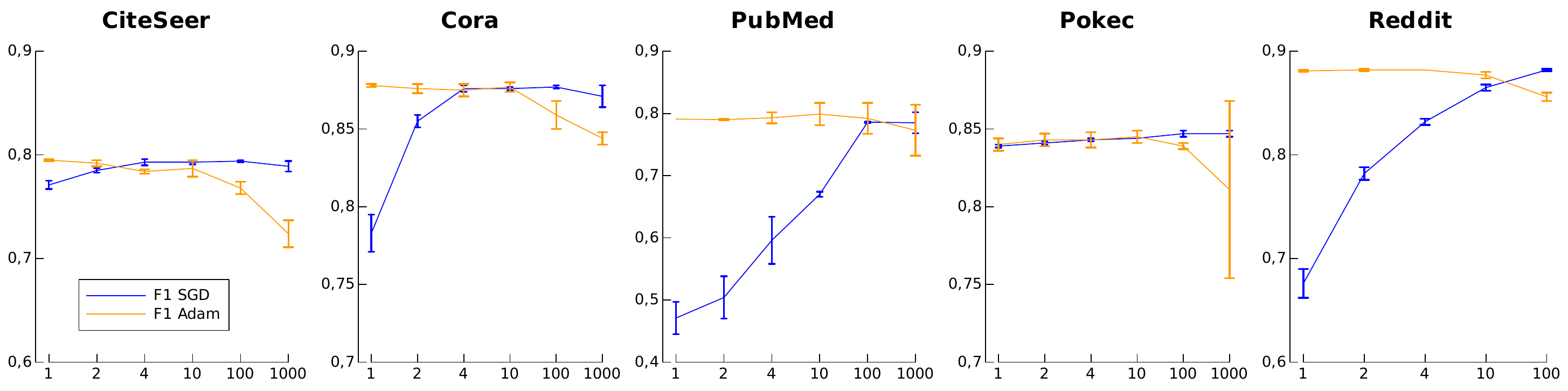}	
	\caption{\label{fig:experiment.e} Experiment C, no DP: $F_1$ wrt. number of subgraphs.}
\end{figure*}

First we highlight main results for the \textbf{non-DP graph splitting} approach.
For all datasets, increasing the number of subgraph divisions yields better results in the SGD setting.
This is especially notable in a case such as for PubMed, where there is an increase from 0.47 with no splits to 0.79 with a splits size of 100. Overall, splitting the graph is shown to be quite effective in a setting where the model may struggle more in the learning process, such as with SGD.
Furthermore, at the higher subgraph split sizes, there is a slight drop-off for 
Adam but not for SGD,
where increasing subgraph split size never improves performance beyond vanilla Adam, as shown in Figure \ref{fig:experiment.e}.

\begin{figure*}
    \centering
	\includegraphics[width=\linewidth]{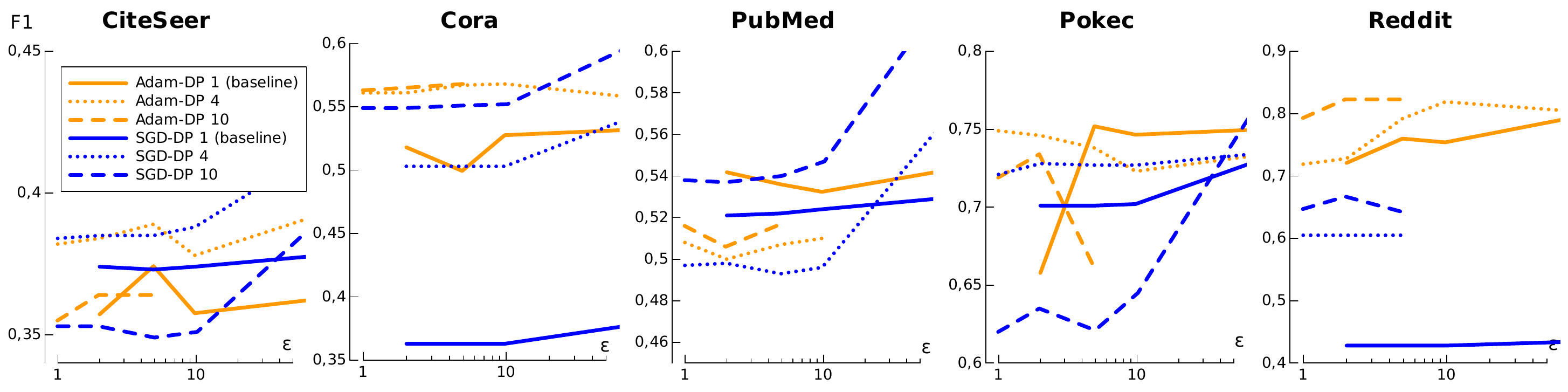}	
	\caption{\label{fig:experiment.f} Experiment C, with DP: $F_1$ with varying number of subgraphs wrt. privacy budget $\varepsilon$.}
\end{figure*}

Figure~\ref{fig:experiment.f} shows the results for the \textbf{graph splitting setting with DP}, varying the privacy budget.
First, increasing the number of \textbf{subgraph splits generally improves results} for SGD-DP (e.g. $0.36 \to 0.55$ for Cora at $\varepsilon=2.0$, with 10 subgraphs vs.\ the full graph, respectively).
We see a difference across datasets, where for instance Reddit shows the best results at 10 splits for all three $\varepsilon$ values, while Citeseer or Pokec do not particularly benefit beyond four splits.

Second, this pattern of improvement is also noticeable in the case of Adam-DP, in contrast to the non-private vanilla Adam results.
This is clearly seen in the Reddit results, where the very best result is with 10 splits with $\varepsilon=1.0$ at an F-score of $0.79$ with Adam-DP, being just $0.09$ points lower than the non-DP version. As in Experiment B, it is notable that Adam-DP does not perform particularly well without using very high learning rates, with less graph splits requiring larger learning rate values.
Overall, the best-performing number of subgraphs seems to be specific to the particular dataset used and can thus be treated as another hyperparameter to optimize on.

Third, with more 
subgraphs
allowing for less noise to be added to obtain a lower $\varepsilon$ value, it is possible to \textbf{reach the strong privacy guarantee} of $\varepsilon=1.0$. For comparison, the randomized response, a DP mechanism extensively used by social scientists for decades \cite{warner1965randomized}, has $\varepsilon \approx 1.1$.
Thus our graph splitting approach not only helps mitigate the difficulties of training with added DP noise, but also allows us to reach stronger privacy guarantees with about the same performance. Without the mini-batching that becomes possible by splitting the graph, it is impossible to carry out a stable computation during the moment accounting process to achieve $\varepsilon=1.0$.
A comparison of the DP setting with graph splitting (using our initial setup of 10 graph splits) and the regular DP setting is summarized in Table~\ref{fig:experiments.ac}.

\paragraph{Summary and take-aways}

We summarize the key observations as follows:
\begin{enumerate}
    \item SGD-DP is fairly robust to noise for these datasets and settings, even at $\varepsilon=1.0$.
    \item Adam-DP works even better than SGD-DP, however it needs to be tuned very carefully, using very high learning rates.
    \item More complex representations are better for the DP setting, showing a smaller performance drop from the non-DP results.
    \item Increasing training data does not necessarily mitigate negative performance effects of DP.
    \item Graph splitting improves both performance and allows for a stronger privacy guarantee of $\varepsilon=1.0$, resolving the mini-batching problem of GCNs in the DP setting.

\end{enumerate}

We provide an additional error analysis in Appendix \ref{sec:hard-cases},
where we show that failed predictions in Reddit and CiteSeer are caused by `hard cases', i.e.\ examples and classes that are consistently misclassified regardless of training data size or privacy budget. Moreover, 
in Appendix \ref{sec:mnist-baselines}
we describe results on the MNIST dataset with varying lot sizes, showing how this hyperparameter affects model results.

\subsection{Pokec Feature Comparison}
\label{sec:pokec-feature-comparison}

As mentioned in Section~\ref{sec:experiment-b}, we perform additional experiments on the Pokec dataset in order to further investigate the hypothesis that input feature complexity has an effect on the degree of performance drop in the DP setting.
We originally noticed this across separate datasets, with Cora, Citeseer and PubMed, using one-hot input features, having a greater performance drop than Reddit and Pokec, which use GloVe and BERT, respectively.
In order to more properly evaluate this under the same conditions, we prepare two additional types of input features for the Pokec dataset, namely Bag of Words (BoWs) and fastText \cite{joulin2016bag}, altogether having three levels of word representations, ranging from the simpler (BoWs) to more complex (BERT).

The BoWs embeddings were prepared by taking the same 20,000 user profiles as in the BERT preprocessing methodology described above. Tokens were split on whitespace, with additional steps such as punctuation removal and lowercasing. In order to reduce the embedding dimensionality, we filtered tokens by frequency in the interval $[15, 15000]$. Each user profile was thus represented with a 9447-dimensional vector of binary values.

For the fastText embeddings, we use a pre-trained model for Slovak from \newcite{grave2018learning}. Using the same set of user profiles, we preprocess the data in the same manner as described by the authors. In order to have one vector per user profile, we average all fastText embeddings for a given user to have a final embedding dimension of 300.

The results of this experiment can be seen in Table~\ref{fig:pokec-features}. We notice that, in line with our hypothesis, the most effective input features in the DP setting are the BERT embeddings, with the smallest performance drop from the non-DP setting (e.g. $0.84 > 0.70$ for SGD-DP at $\varepsilon=2.0$). Interestingly, the best method overall without DP is with the BoWs representation. One explanation for this is that a lot of slang vocabulary and unusual tokens are used in the social network data, which a fastText or BERT model may struggle with more, while BoWs would simply treat them equally as any other token in the vocabulary. As expected, the BoWs embeddings have a larger drop when trained with DP ($0.88 > 0.62$ for SGD and SGD-DP with $\varepsilon=2.0$, respectively). The fastText results show the lowest performance both in the non-DP and DP settings, possibly due to the model struggling to maintain useful representations after averaging many token vectors for a user, which a more powerful model such as mBERT has an easier time with.
Our original hypothesis is thus verified that more sophisticated input features such as BERT would show a smaller performance drop in the DP setting, compared to simpler representations such as BoWs, with this effect shown in an `apples-to-apples' setting on the same dataset.

\begin{table}
    \centering

		\begin{tabular}{ll|lll}
			\multicolumn{2}{c}{\textbf{Non-DP} $F_1$ scores} & \multicolumn{3}{c}{\textbf{DP} $F_1$ scores} \\ \hline
			SGD	&Adam	&$\varepsilon$	&SGD	&Adam	\\ \midrule
			\multicolumn{2}{l|}{\textbf{BoWs}} &2	&0.62	&0.63	\\
            0.88 & 0.87	&5	&0.62	&0.61	\\
            &	&10	&0.62	&0.61	\\
            &	&137	&0.63	&0.63	\\ \hline
			\multicolumn{2}{l|}{\textbf{fastText}} &2	&0.59	&0.63	\\
            0.71 & 0.73	&5	&0.59	&0.57	\\
             &	&10	&0.59	&0.57	\\
             &	&137	&0.61	&0.60	\\ \hline
			\multicolumn{2}{l|}{\textbf{BERT}} &2	&0.70	&0.66	\\
            0.84 & 0.84	&5	&0.70	&0.75	\\
             &	&10	&0.70	&0.75	\\
             &	&137	&0.74	&0.75	\\
		\end{tabular}
	    \caption{\label{tab:pokec-feature-experiments} $F_1$ scores for the Pokec dataset, comparing different input feature representations and privacy budgets.}
	\label{fig:pokec-features}
\end{table}

\section{Conclusion}

We have explored differentially-private training for GCNs, showing the nature of the privacy-utility trade-off.
We show that an expected drop in results for the DP models can be mitigated by graph partitioning, utilizing Adam-DP, as well as having more complexity in the input representations.
Our approach achieves strong privacy guarantees of $\varepsilon = 1.0$, yet reaching up to 87\% and 90\% of non-private $F_1$ scores for Pokec and Reddit datasets, respectively.
An interesting line of future work is to explore further options for graph splitting that may take advantage of the graph structure instead of uniformed random splitting.
By adapting global DP to a challenging class of deep learning networks, we are thus a step closer to flexible and effective privacy-preserving NLP.

\section*{Acknowledgments}
This research work has been funded by the German Federal Ministry of Education and Research and the Hessian Ministry of Higher Education, Research, Science and the Arts within their joint support of the National Research Center for Applied Cybersecurity ATHENE.
Calculations were conducted on the Lichtenberg high performance computer of the TU Darmstadt.

\section{Bibliographical References}
\label{reference}

\bibliographystyle{lrec2022-bib}
\bibliography{bibliography}

\clearpage

\appendix

\section{Hyperparameter Configuration}
\label{sec:hyperparams}

Our GCN model consists of 2 layers, with ReLU non-linearity, a hidden size of 32 and dropout of 50\%, trained with a learning rate of 0.01 (apart from Adam-DP, which required far higher learning rates, as mentioned below). We found that early stopping the model works better for the non-DP implementations, where we used a patience of 20 epochs. We did not use early stopping for the DP configuration, which shows better results without it. For all SGD runs we used a maximum of 2000 epochs, while for Adam we used 500.

Importantly, for Adam-DP we noticed that more moderate learning rate values such as 0.01 were insufficient and led to far lower performance. We therefore optimized this at several values in the interval from 0.1 to 100, with some datasets and graph split values requiring learning rates as low as 0.1 (e.g. most datasets with 100 graph splits), while in other cases requiring 50 or 100 (e.g. Reddit for most graph split values).

Due to the smaller amount of epochs for Adam, it is possible to add less noise to achieve a lower $\varepsilon$ value. Table~\ref{tab:epsilon-to-noise} shows the mapping from noise values used for each optimizer to the corresponding $\varepsilon$ in the full graph setting.
The runtimes for our experiments reach up to 1 hour for the larger configurations on an NVIDIA A100 GPU.

\begin{table}[h!]
\centering
\caption{\label{tab:epsilon-to-noise} $\varepsilon$ values from experiment B, with the corresponding noise values added to the gradient for each optimizer.}
\begin{tabular}{l|rr}
\textbf{$\varepsilon$}	&\textbf{Noise-SGD}	&\textbf{Noise-Adam}\\ \midrule
136.51	&4	&2\\
9.75	&26	&13\\
4.91	&48	&24\\
2.00	&112	&56\\
\end{tabular}
\end{table}

Finally, regarding hyperparameter optimization on the validation set in the DP setting, \newcite{Abadi.et.al.2016.SIGSAC} mention that, when optimizing on a very high number of parameter settings (e.g. in the thousands), this would additionally take up a moderate privacy budget (e.g. $\varepsilon=4$ if they had used 6,700 hyperparameters). For our experiments, this number of hyperparameters is comparatively minimal and would be well within our privacy bounds.

\section{Are `hard' examples consistent between private and non-private models?}
\label{sec:hard-cases}

\begin{figure}
\includegraphics[width=0.48\textwidth]{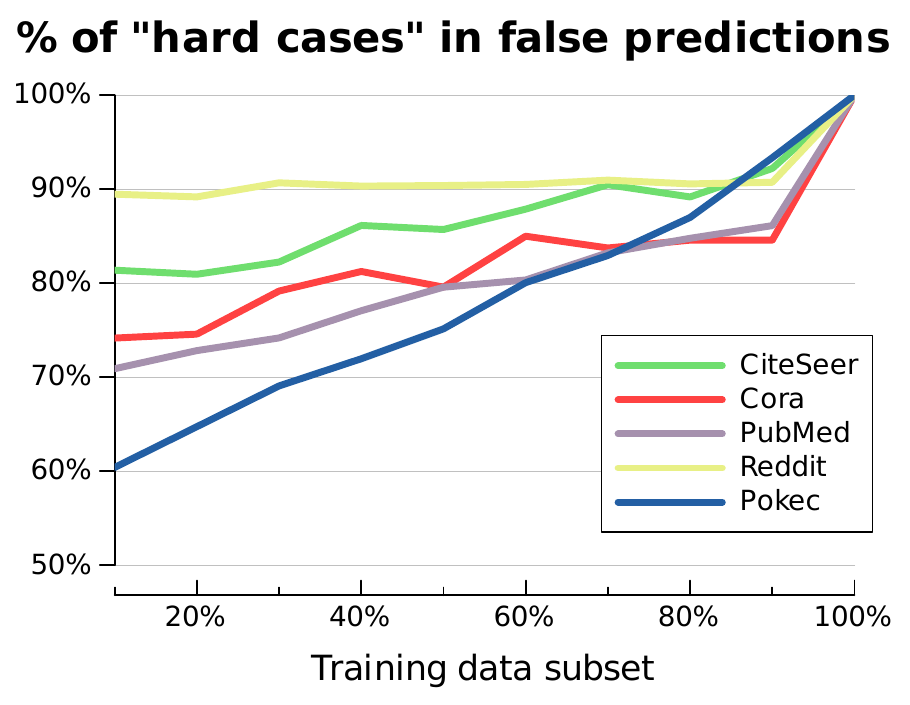}
\caption{\label{fig:hard.cases} Hard cases in non-DP.}
\end{figure}

\begin{figure}
\includegraphics[width=0.48\textwidth]{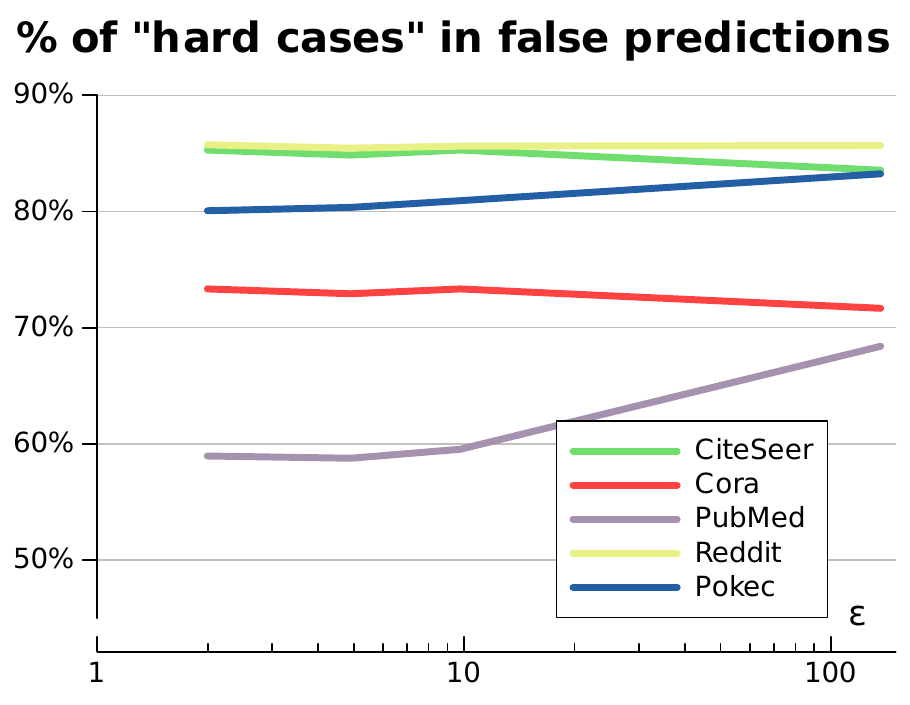}
\caption{\label{fig:hard-cases-analysis-dp} Hard cases analysis DP.}
\end{figure}

To look further into the nature of errors for experiments A and B, we evaluate the `hard cases'. These are cases that the model has an incorrect prediction for with the maximum data size and non-private implementation (the first set of results of experiment A). For the experiment A learning curves, we take the errors for every setting of the experiment (10\% training data, 20\%, and so forth) and calculate the intersection of those errors with that of the `hard cases' from the baseline implementation. This intersection is then normalized by the original number of hard cases to obtain a percentage value. The results for experiment A can be seen in Figure~\ref{fig:hard.cases}. We perform the same procedure for experiment B with different noise values for the SGD-DP setting, as seen in Figure~\ref{fig:hard-cases-analysis-dp}. This provides a look into how the nature of errors differs among these different settings, whether they stay constant or become more random as we decrease the training size or increase DP noise.

Regarding the errors for experiment B, we can see a strong contrast between datasets such as Reddit and PubMed. For the latter, the more noise we add as $\varepsilon$ decreases, the more random the errors become. In the case of Reddit, however, we see that even if we add more noise, it still fails on the same hard cases. This means that there are hard aspects of the data that remain constant throughout. For instance, out of all the different classes, some may be particularly difficult for the model.

Although the raw data for Reddit does not have references to the original class names and input texts, we can still take a look into these classes numerically and see which ones are the most difficult in the confusion matrix. In the baseline non-DP model, we notice that many classes are consistently predicted incorrectly. For example, class 10 is predicted 93\% of the time to be class 39. Class 18 is never predicted to be correct, but 95\% of the time predicted to be class 9. Class 21 is predicted as class 16 83\% of the time, and so forth. This model therefore mixes up many of these classes with considerable confidence.

Comparing this with the confusion matrix for the differentially private implementation at an $\varepsilon$ value of 2, we can see that the results incorrectly predict these same classes as well, but the predictions are more spread out. Whereas the non-private model seems to be very certain in its incorrect prediction, mistaking one class for another, the private model is less certain and predicts a variety of incorrect classes for the target class.

For the analysis of the hard cases of experiment A in Figure~\ref{fig:hard.cases}, we can see some of the same patterns as above, for instance between PubMed and Reddit. Even if the training size is decreased, the model trained on Reddit still makes the same types of errors throughout. In contrast, as training size is decreased for PubMed, the model makes more and more random errors. The main difference between the hard cases of the two experiments is that, apart from Reddit, here we can see that for all other datasets the errors become more random as we decrease training size. For example, Cora goes down from 85\% of hard cases at 90\% training data to 74\% at 10\% training data. In the case of experiment B, they stay about the same, for instance Cora retains just over 70\% of the hard cases for all noise values.

Overall, while we see some parallels between the hard cases for experiments A and B with respect to patterns of individual datasets such as Reddit and PubMed, the general trend of more and more distinct errors that is seen for the majority of datasets with less training size in experiment A is not the same in experiment B, staying mostly constant across different noise values for the latter. The idea that the nature of errors for DP noise and less training data being the same is thus not always the case, meaning that simply increasing training size may not necessarily mitigate the effects of DP noise.    

\section{MNIST Baselines}
\label{sec:mnist-baselines}

\begin{table}
\centering
\caption{\label{tab:mnist-baselines} Results on the MNIST dataset with varying lot sizes and noise values.}
\begin{tabular}{lll|cc}
\textbf{Lot Size}	&\textbf{Noise} & \textbf{$\varepsilon$}	&\textbf{$F_1$} & \textbf{Std.}\\ \midrule
600	& 4	& 1.26 & 0.90 & 0.02\\
6,000	& 4	& 4.24 & 0.84 & 0.01\\
60,000	& 4	& 15.13 & 0.45 & 0.04\\
60,000	& 50	& 0.98 & 0.39 & 0.15\\
60,000	& 100	& 0.50 & 0.10 & 0.01\\
\end{tabular}
\end{table}

Table~\ref{tab:mnist-baselines} shows results on the MNIST dataset with different lot sizes and noise values, keeping lot and batch sizes the same. We use a simple feed-forward neural network with a hidden size of 512, dropout of 50\%, SGD optimizer, and a maximum of 2000 epochs with early stopping of patience 20, with other hyperparameters such as learning rate being the same as above.
We note that the configuration in the first row with lot size of 600 and noise 4 is the same as described by \newcite{Abadi.et.al.2016.SIGSAC} in their application of the moments accountant, reaching the same $\varepsilon$ value of 1.2586.

We can see some important patterns in these results that relate to our main results from the GCN experiments. Maintaining a constant noise of 4, as we increase the lot size, not only does the $\varepsilon$ value increase, but we see a dramatic drop in $F_1$ score, especially for a lot size of 60,000, being the full training set. If we try to increase the noise and maintain that 60,000 lot size, while we are able to lower the $\varepsilon$ value below 1, the $F_1$ score continues to drop dramatically, going down to 0.1010 with a noise value of 100.

Hence, the current MNIST results further show the benefits of applying the graph splitting methodology on large one-graph datasets. By splitting the graph, we are able to utilize batches and lots of smaller sizes, allowing to add less noise to reach a lower epsilon value, ultimately retaining a higher $F_1$ score in the DP setting.

The four English datasets adapted from the previous work are only available in their encoded form. For the citation networks, each document is represented by a bag-of-words encoding.

The Reddit dataset combines GloVe vectors \cite{pennington-etal-2014-glove} averaged over the post and its comments. Only the Pokec dataset is available as raw texts, so we opted for multilingual BERT \cite{devlin2018bert} and averaged all contextualized word embeddings over each users' textual attributes.\footnote{Sentence-BERT \cite{reimers2019sentence} resulted in lower performance. Users fill in the attributes such that the text resembles a list of keywords rather than actual discourse. } The variety of languages, sizes, and different input encoding allows us to compare non-private and private GCNs under different conditions.
Table~\ref{tab1:data-summary} summarizes data sizes and number of classes.

\begin{table}[h!]
	\centering
	\caption{\label{tab1:data-summary} Dataset statistics; size is number of nodes.}
	\begin{tabular}{lr|rr}
		\textbf{Dataset}	& \textbf{Classes}	& \textbf{Test size}	&\textbf{Training size}		\\ \midrule
		CiteSeer	&6	&1,000	&1,827	\\
		Cora	&7	&1,000	&1,208		\\
		PubMed	&3	&1,000	&18,217	\\
		Pokec	&2 	&2,000 	&16,000 	\\
		Reddit	&41	&5,643	&15,252	\\
	\end{tabular}
\end{table}

\section{Further details on Pokec dataset pre-processing}
\label{sec:pokec-preprocessing}

In order to prepare the binary classification task for the Pokec dataset, the original graph consisting of 1,632,803 nodes and 30,622,564 edges is sub-sampled to only include users that filled out the `pets' column and had either cats or dogs as their preference, discarding entries with multiple preferences. For each pet type, users were reordered based on percent completion of their profiles, such that users with most of the information were retained.

For each of the two classes, the top 10,000 users are taken, with the final graph consisting of 20,000 nodes and 32,782 edges. The data was split into 80\% training, 10\% validation and 10\% test partitions.

The textual representations themselves were prepared with `bert-multilingual-cased' from Huggingface transformers,\footnote{https://github.com/huggingface/transformers} converting each attribute of user input in Slovak to BERT embeddings with the provided tokenizer for the same model.
Embeddings are taken from the last hidden layer of the model, with dimension size 768. The average over all tokens is taken for a given column of user information, with 49 out of the 59 original columns retained. The remaining 10 are left out due to containing less relevant information for textual analysis, such as a user's last login time.
To further simplify input representations for the model, the average is taken over all columns for a user, resulting in a final vector representation of dimension 768 for each node in the graph.

\section{Moments Accountant in Detail}
\label{sec:moments.accountant}

SGD-DP introduces two features, namely (1) a reverse computation of the privacy budget, and (2) tighter bounds on the composition of multiple queries.
First, a common DP methodology is to pre-determine the privacy budget ($\varepsilon, \delta$) and add random noise according to these parameters. In contrast, SGD-DP does the opposite: Given a pre-defined amount of noise (hyper-parameter of the algorithm), the privacy budget ($\varepsilon, \delta$) is computed retrospectively. Second, generally in DP, with multiple executions of a `query' (i.e.\ a single gradient computation in SGD), we can simply sum up the $\varepsilon, \delta$ values associated with each query, such that for $k$ queries with privacy budget $(\varepsilon, \delta)$, the overall algorithm is $(k \varepsilon, k \delta)$-DP. However, this naive composition leads to a very large privacy budget as it assumes that each query used up the maximum given privacy budget.

The simplest bound on a continuous random variable $Z$, the Markov inequality, takes into account the expectation $\mathbb{E}[Z]$, such that for $\varepsilon \in \mathbb{R}^+$:

\begin{equation}
\label{eq:markov.ineq}
\Pr[Z \geq \varepsilon] \leq \frac{\mathbb{E}[Z]}{\varepsilon}
\end{equation}

Using the Chernoff bound, a variant of the Markov inequality, on the privacy loss $Z$ treated as a random variable 
(Eq. 2),
we obtain the following formulation by multiplying Eq.~\ref{eq:markov.ineq} by $\lambda \in \mathbb{R}$ and exponentiating:

\begin{equation}
\label{eq:chernoff-unsimplified1}
\Pr[\exp(\lambda Z) \geq \exp(\lambda \varepsilon)] \leq \frac{ \mathbb{E} [\exp(\lambda Z)]}{\exp(\lambda \varepsilon)}
\end{equation}
where $\mathbb{E} [\exp(\lambda Z)]$ is also known as the moment-generating function.

The overall privacy loss $Z$ is composed of a sequence of consecutive randomized algorithms $X_1, \dots, X_k$. Since all $X_i$ are independent, the numerator in Eq.~\ref{eq:chernoff-unsimplified1} becomes a product of all $\mathbb{E}[\exp(\lambda X_i)]$. Converting to log form and simplifying, we obtain

\begin{equation}
\label{eq:chernoff-simplified}
\Pr[Z \geq \varepsilon] \leq \exp \left( \sum_i \ln \mathbb{E}[\exp(\lambda X_i)] - \lambda \varepsilon \right)
\end{equation}

Note the moment generating function inside the logarithmic expression. Since the above bound is valid for any moment of the privacy loss random variable, we can go through several moments and find the one that gives us the lowest bound.

Since the left-hand side of Eq.~\ref{eq:chernoff-simplified} is by definition the $\delta$ value, the overall mechanism is ($\varepsilon, \delta$)-DP for $\delta = \exp(\sum_i \ln \mathbb{E}[\exp(\lambda X_i)] - \lambda \varepsilon)$. The corresponding $\varepsilon$ value can be found by modifying \ref{eq:chernoff-simplified}:

\begin{equation}
\label{eq:chernoff-epsilon}
\varepsilon = \frac{\sum_i \ln \mathbb{E}[\exp(\lambda X_i)] - \ln \delta}{\lambda}
\end{equation}

The overall SGD-DP algorithm, given the right noise scale $\sigma$ and a clipping threshold $C$, is thus shown to be $(O(q\varepsilon\sqrt{T}), \delta)$-differentially private using this accounting method, with $q$ representing the ratio $\frac{L}{N}$ between the lot size $L$ and dataset size $N$, and $T$ being the total number of training steps. See \cite{Abadi.et.al.2016.SIGSAC} for further details.

\end{document}